\documentclass[conference]{IEEEtran}
\IEEEoverridecommandlockouts
\usepackage{cite}
\usepackage{amsmath,amssymb,amsfonts}
\usepackage{algorithmic}
\usepackage{graphicx}
\usepackage{textcomp}
\usepackage{xcolor}

\usepackage[hyphens]{url}
\usepackage[hidelinks]{hyperref}
\hypersetup{breaklinks=true}
\usepackage{soul}
\usepackage{float}
\usepackage{multirow}
\usepackage{tabularx}
\usepackage{tabulary}
\usepackage{booktabs}
\usepackage{verbatim}
\usepackage[np, autolanguage]{numprint}
\usepackage{subfloat}

\usepackage[all=normal,charwidths]{savetrees}

\ifCLASSOPTIONcompsoc
  \usepackage[caption=false,font=normalsize,labelfont=sf,textfont=sf]{subfig}
\else
  \usepackage[caption=false,font=footnotesize]{subfig}
\fi

\newcommand{\lmttfont}{\fontfamily{lmtt}\selectfont}

\newcommand{\toolname}[1]{\emph{ProFIPy}}
\newcommand{\pydsl}[1]{\emph{PyDSL}}

\usepackage{listings}
  \lstdefinelanguage{diff}{
    basicstyle=\scriptsize\ttfamily,
    frame=single,
    breaklines=true,
  }

\def\BibTeX{{\rm B\kern-.05em{\sc i\kern-.025em b}\kern-.08em
    T\kern-.1667em\lower.7ex\hbox{E}\kern-.125emX}}
\begin{document}

\title{\toolname{}: Programmable Software Fault Injection as-a-Service}

\author{
\IEEEauthorblockN{Domenico Cotroneo, Luigi De Simone, Pietro Liguori, Roberto Natella}
\IEEEauthorblockA{\textit{Universit\`a degli Studi di Napoli Federico II, Italy} \\  
\{cotroneo, luigi.desimone, pietro.liguori, roberto.natella\}@unina.it}
}

\maketitle

\thispagestyle{plain}
\pagestyle{plain}

\begin{abstract}
In this paper, we present a new fault injection tool (\toolname{}) for Python software. The tool is designed to be \emph{programmable}, in order to enable users to specify their software fault model, using a \emph{domain-specific language} (DSL) for fault injection. Moreover, to achieve better usability, \toolname{} is provided as \emph{software-as-a-service} and supports the user through the configuration of the faultload and workload, failure data analysis, and full automation of the experiments using container-based virtualization and parallelization.
\end{abstract}

\begin{IEEEkeywords}
Software Fault Injection, Python, Software-as-a-Service, Bug Pattern
\end{IEEEkeywords}

\section{Introduction}
\label{sec:introduction}


Fault injection is a key technique for assessing fault-tolerant systems, ranging from embedded and mobile systems \cite{hsueh1997techniques} to distributed systems \cite{joshi2011prefail}. To perform a fault injection campaign, it is important to define a \emph{fault model}, which describes the faults to be emulated in the experiments. The fault model entails the definition of three main aspects, namely \emph{what} to inject (i.e., which kind of fault), \emph{when} to inject (i.e., the timing of the injection), and \emph{where} to inject (i.e., the part of the system targeted by the injection) \cite{chillarege1996:generation-error-set,madeira2000:emulation-swifi,johansson2005error,lanzaro2014empirical, cotroneo2012experimental}. The \emph{what} can be represented by bit-flips \cite{hsueh1997techniques}; program exceptions for amplifying unit- and integration-tests \cite{adamsen2015systematic,jiang2019fuzzing}; node crashes, network partitions and latency for networked and distributed systems \cite{joshi2011prefail,gunawi2011fate}. The \emph{when} and \emph{where} to inject are sampled from a (large) space of possibilities across time and program locations.

The problem of defining a fault model becomes more difficult when injecting \emph{software faults} (i.e., design and/or programming defects \cite{avizienis2004:basicconcepts}), since they depend on a variety of technical and organizational factors, including the programming language, the software development process, the maturity of the system, the expertise of developers, and the application domain \cite{huang2012taxonomy,huang2017human}.
%
%
Despite the variability of software faults across systems, the existing software fault injection tools are based on a predefined, fixed software fault model, that cannot be easily customized by users. Most of the existing tools adopt the \emph{Orthogonal Defect Classification} (ODC), proposed in the '90s (e.g., bugs in initialization, algorithm, interfaces, etc.), or derived the fault model from bug samples of third-party open-source and commercial projects \cite{chillarege1996:generation-error-set,duraes2006:emulationswfaults}. 

We believe that a modern software fault injection tool has to be able to modify the fault model for the following reasons. 
First, a typical necessity in industry, which arises when a critical failure occurs, is to introduce regression tests against the fault that caused the failure, to assure that the same failure cannot occur again \cite{yuan2014simple}.  
Second, to preserve the efficiency of the fault injection campaign, it is important to avoid injecting bugs that are unlikely to affect a system; e.g., some classes of faults may be prevented by testing and static analysis policies adopted by the company \cite{bessey2010few}. 
Third, as the scale and the complexity of systems increase, the need for a more sophisticated fault model grows. For instance, modern distributed systems, such as cloud applications, have to integrate a variety of components, including third-party and open-source ones, and they have to deal with high volumes of traffic. 
For these systems, the user needs to inject more variants of design/programming defects than those reported in the literature, including performance bottlenecks, resource management issues, lack of interoperability between components, security issues, failed updates, etc., and to adapt these faults to their projects. 
In general, the potential users of software fault injection want to tune the fault model so that it reflects their experience and expectations about failures. All these use cases require a greater degree of control over the fault model than what provided by existing fault injection tools.

In this paper, we present a new fault injection tool (\toolname{}) designed to be \emph{programmable}, enabling users to add and to customize a software fault model. 
By using our tool, users can specify new software fault models using a \emph{domain-specific language} (DSL) for fault injection. The tool compiles the specification into an automatically-generated fault injector. Finally, the generated fault injector is applied to the software-under-test to generate fault-injected versions and to execute experiments. 
To achieve better usability, \toolname{} is provided as \emph{software-as-a-service}, and includes a workflow for configuring the faultload and the workload to i) fully automate the execution of experiments using container-based virtualization and parallelization, and to ii) perform failure data analysis. The tool has been designed for the popular Python language, which has recently arisen as one of the most widespread languages (e.g., among the GitHub and StackOverflow communities \cite{octoverse,stackoverflowsurvey}), and has found applications in several areas such as systems software (e.g., the OpenStack cloud platform is one of the largest projects in Python \cite{OpenStackUsers,OpenStackProducts}), enterprise and web applications and data science \cite{pythonapps}.
We present \toolname{} in the context of a Python project, by performing three fault injection campaigns in which we define three different faultloads.

In the following, Section~\ref{sec:related} discusses related work; Section~\ref{sec:pycast_dsl} presents a new domain-specific language; Section~\ref{sec:pycast_overview} describes the workflow of the tool; Section~\ref{sec:case_study} shows the application of \toolname{} on a Python project; Section~\ref{sec:conclusion} concludes the paper.

\section{Related Work}
\label{sec:related}

The idea of software fault modeling for fault injection purposes was initially investigated by Chillarege et al. \cite{chillarege1988understanding}, who analyzed a dataset of failures of IBM OS and DBMS products at users' sites \cite{sullivan1991:defectstudy,sullivan1992comparison}, to identify recurring patterns in the faults that caused them, and to inject the same patterns by corrupting program data and code, e.g., as in the \emph{FINE} tool \cite{kao1993:fine}. 
In the same period, they also introduced the \emph{Orthogonal Defect Classification} (ODC) \cite{chillarege1991:growth,chillarege1992:odc}, where one the goals was to classify software fault data into orthogonal categories, including \emph{Initialization}, \emph{Algorithm}, \emph{Interface}, \emph{Checking}, and \emph{Synchronization} defects. Christmansson and Chillarege \cite{chillarege1996:generation-error-set} proposed to inject software faults by following the statistical distribution of OS faults across these categories, such that the injected faults are representative of faults experienced by the users of the OS in the field.  Similarly, Chen and colleagues \cite{ng1996comparing,ng2001design} defined a software fault model for OSes based on data for the IBM MVS and Tandem GUARDIAN90 OS products \cite{sullivan1991:defectstudy,iyer1993:faultstandem}, and used this fault model to emulate realistic OS and DBMS crashes, to assess crash recovery mechanisms. 
This fault model was later merged in the well-known fault injection tool of the \emph{Nooks} project \cite{swift2006recovering}. 

The work on the \emph{G-SWFIT} fault injection technique by Madeira and colleagues \cite{duraes2002:educatedmutations,duraes2006:emulationswfaults} aimed to define a \emph{generic} software fault model (i.e., not tailored for a specific system) that could go beyond specific OS and DBMS products, and that could be used for injecting faults even without any field failure data for the specific system under testing. To define such a generic fault model, they analyzed a sample of bugs in several open-source projects in C \cite{duraes2002:educatedmutations,duraes2006:emulationswfaults} and Java \cite{basso2009java,sanches2011:jswfit}, and looked for bug-fixes (e.g., program elements that were changed to fix the bug, such as new assignments, control flow constructs, function calls, etc.) which were recurring more than the norm, and which occurred consistently across all of the projects. Based on this analysis, they defined a software fault model with 13 fault types, covering 60\% of the sample of bugs in the open-source projects \cite{duraes2006:emulationswfaults}. This fault model was used in several other tools, including \emph{SAFE} \cite{cotroneo2013fault}, \emph{HSFI} \cite{van2016hsfi}, and \emph{FastFI} \cite{schwahn2018fastfi}. However, these tools focus on a fixed software fault model, with no ability to customize the injected faults according to the specific needs of a project or company.

Winter et al. \cite{winter2011impact} and Giuffrida et al. \cite{giuffrida2013edfi} showed that implementing a new fault model in a tool takes both significant programming effort, e.g., in terms of SLOC and other metrics, and considerable expertise in program analysis and transformation, e.g., to implement a software fault injection tool using the \emph{LLVM} compiler suite, which are not affordable for the average user of a fault injection tool.

Some tools provide a limited ability to customize the fault model with a lower effort: among them, the \emph{FIDLFI} tool \cite{aliabadi2016fidl} provides the user with a configuration language to control the \emph{trigger} of fault injection (i.e., instructions and paths that trigger the injection), \emph{target} (i.e., instruction source and destination registers to inject), and \emph{action} (e.g., corruption, freeze, delay, etc.). The \emph{FAIL-FCI} tool \cite{hoarau2007fail} provides a fault injection language tailored for grid systems, which specifies protocol states and nodes to inject (e.g., node crashes). \emph{PreFail} \cite{joshi2011prefail} and \emph{FATE} \cite{gunawi2011fate}, which inject crashes and I/O API errors, allow the user to write \emph{policies} in Python to select the location and timing of potential injections by considering the allocation of processes across nodes and racks (e.g., network partitions between different racks), and the coverage of injectable points in the software-under-test. \emph{LFI} \cite{marinescu2009lfi}, which injects errors at C library calls, allows the user to configure what functions and error codes should be injected, and when to trigger the injection (e.g., when a specific function appears in the stack frame) using an XML configuration file. The commercial tools \emph{QA Systems Cantata} \cite{qasystemscantata} and \emph{Razorcat TESSY} \cite{razorcattessy} provide user-friendly GUIs to select a source-code statement to inject, similarly to breakpoints in a GUI debugger.

It is important to note that these tools do not support rich software fault models as in \emph{G-SWFIT} and derivatives, as they only provide limited control on \emph{what} to inject, e.g., they focus on API and library calls, register accesses, nodes, etc., but do not allow to create new fault types for injecting arbitrary changes to the software. The proposed \toolname{} tool provides a new language to gain a higher degree of control, where the user can specify transformation rules about which parts of the program to inject, in terms of program elements (e.g., assignments, expressions, control flow directives, and combinations of thereof), and how to transform these program elements into faulty ones.

\section{Fault Injection Domain-Specific Language}
\label{sec:pycast_dsl}

The \toolname{} allows the user to enter a \emph{bug specification} using a high-level and easy-to-use DSL language, which is close to the Python language. The bug specification describes how the source code of the program should be transformed to introduce a software bug. It consists of two parts:

\begin{itemize}
\item \textbf{Code pattern}: a description of which parts of the program should be fault-injected. The fault injection tool parses the source code of the software and will generate a fault for every match of the code pattern.
    
\item \textbf{Code replacement}: a description of the code that should be injected, which will replace the original source code that matched the code pattern.

\end{itemize}

The code pattern describes a combination of program entities (variables, expressions, blocks, control flow constructs, etc.) that will be searched for in the software-under-injection. The code pattern can either consist of a Python snippet of code; or, it can be a mix of Python code and DSL directives. In the former case, \toolname{} will look for \emph{exact} matches between the Python snippet in the code pattern and the Python code in the software-under-injection. In the latter case, the DSL directives will make the pattern to match several different variants of the Python snippet of code. Similarly, the code replacement can either be Python-only code, i.e., the injector will insert a fixed snippet of buggy code; or, it can contain a mix of Python and DSL directives, i.e., the injected buggy code can vary depending on what matched the code pattern.

\begin{figure}[!htb]
       \centering
       \subfloat[Missing function call fault (MFC).]{\label{fig:dsl_example_1}\includegraphics[scale=0.32]{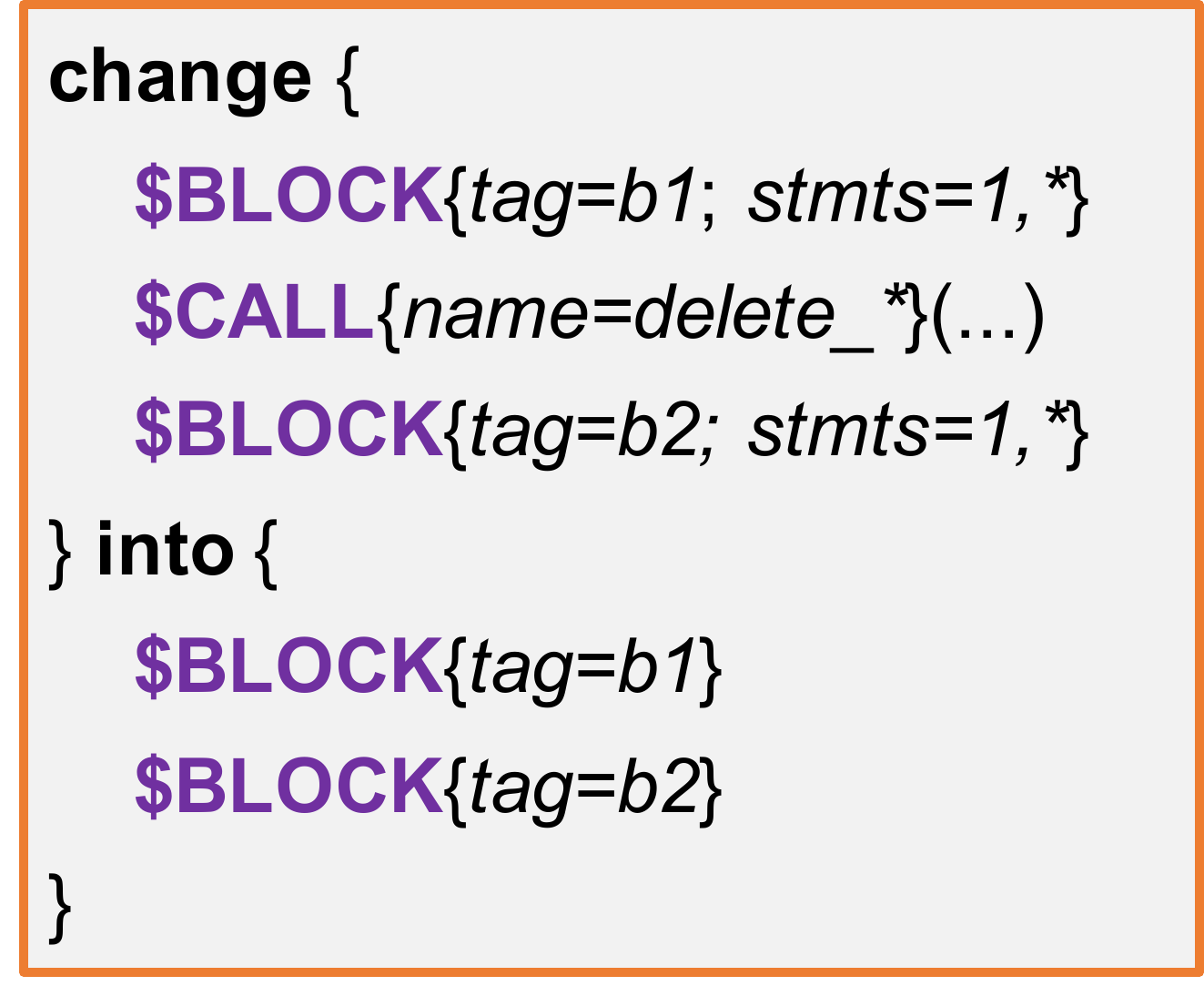}}%
       \qquad%
       \subfloat[Missing IF construct with statements (MIFS) fault.]{\label{fig:dsl_example_2}\includegraphics[scale=0.32]{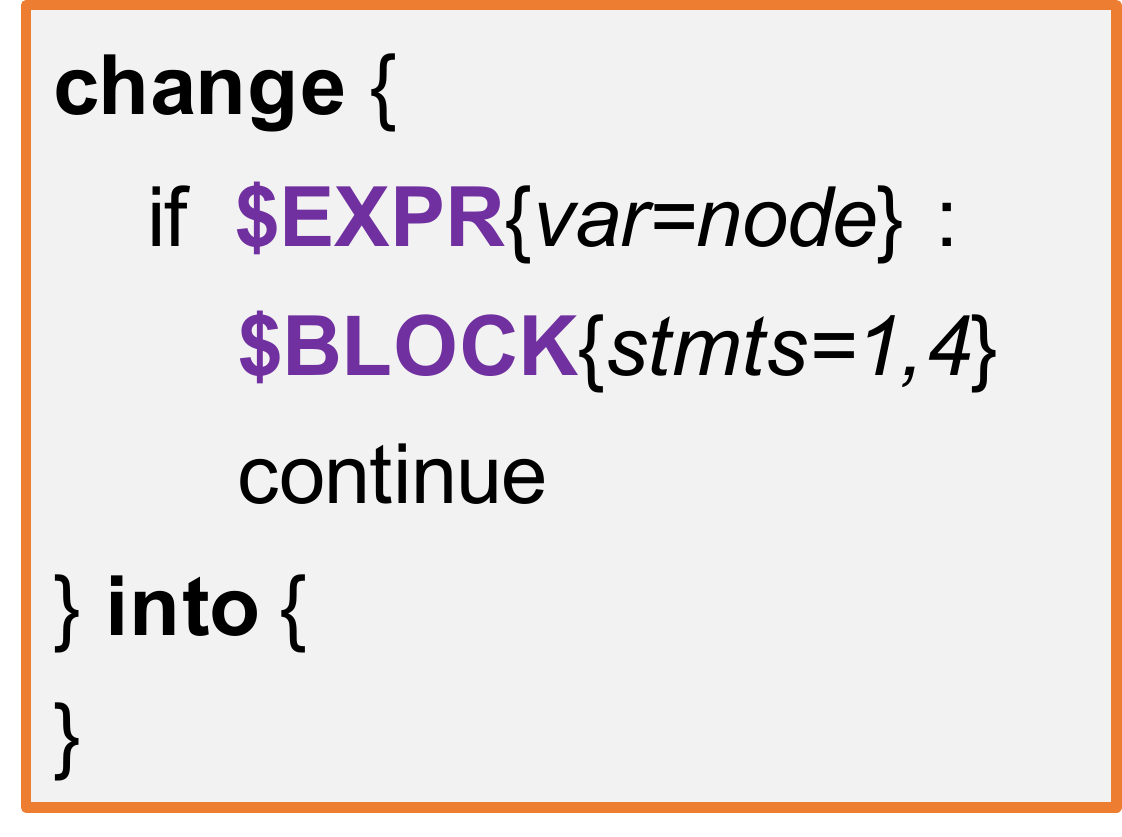}}%
       
       \subfloat[Wrong parameter in function call (WPF) fault.]{\label{fig:dsl_example_3}\includegraphics[scale=0.32]{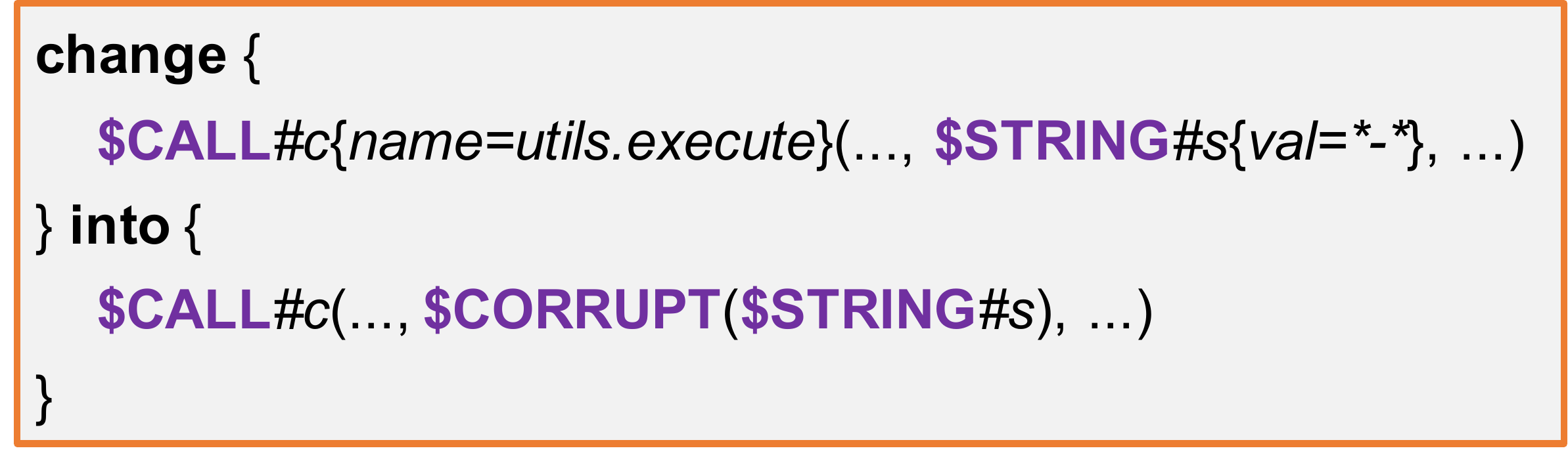}}

        \caption{Examples of fault specifications.}
        \label{fig:dsl_examples}
       \vspace{-5pt}
\end{figure}

\figurename{}~\ref{fig:dsl_examples} shows three examples of bug specifications. These specifications inject three fault types from G-SWFIT \cite{duraes2006:emulationswfaults}: the omission of a function call (MFC); the omission of a small block of statements surrounded by an IF construct (MIFS); and a wrong parameter in input to a function call (WPF). Differing from the G-SWFIT technique, we modified the definition of the fault types, to point out the features of the DSL language, and to emulate more accurately some of the bugs that we found in the OpenStack project \cite{cotroneo2019bad,cotroneo2019analyzing}. 

The MFC fault type from G-SWFIT looks for function calls in the software-under-injection, where there is no return value from the function call, or where the return value is ignored by the caller \cite{duraes2006:emulationswfaults}. By targeting this kind of function calls, the injector can emulate a function call omission by removing these function call statements, and yet to obtain a syntactically-correct program, as the removal does not break any dependency with the rest of the program. Moreover, the G-SWFIT study \cite{duraes2006:emulationswfaults} recommended that the function call should only be removed when the function call is not the only statement in its block, to better reflect the real bugs from open-source projects that were analyzed in that study. 

In \figurename{}~\ref{fig:dsl_example_1}, the code pattern (i.e., the \emph{change \{ $\ldots$ \}} part of the specification) looks for any function or method call, by using the {\lmttfont\$CALL} directive of the DSL. 
The {\lmttfont\{name=delete\_*\}} syntax after {\lmttfont\$CALL} means that we are targeting calls where the function name starts with ``\emph{delete\_}'' string, in order to inject faults in calls to the OpenStack Neutron APIs {\lmttfont delete\_port}, {\lmttfont delete\_subnet}, {\lmttfont delete\_network}, etc. This is an example of how a user may want to customize fault injection according to domain knowledge: these APIs are prone to omissions (e.g., the Neutron bug \#1028174 \cite{bugopenstack1}), and users may want to simulate these faults to assess solutions for resource leak detection. 
The rest of the specification implements the rules of the MFC fault type. {\lmttfont \$CALL} only matches statements where the function or method call is the outermost part of the statement: thus, a statement like {\lmttfont x = mycall()}, where the assignment is the outermost expression, would not match the code pattern of \figurename{}~\ref{fig:dsl_example_1}. The \emph{(\ldots)} syntax means that we are targeting function calls with any number of input parameters (zero, one, or more). The directives {\lmttfont\$BLOCK} directives require that the function call must be both preceded and followed by one or more statements. 
Finally, the code replacement (i.e., the \emph{into \{ $\ldots$ \}} part of the specification) means that we want to transform the matched code by replacing it only with the blocks that precede and follow the function call. The {\lmttfont \{tag=...\}} syntax after {\lmttfont \$BLOCK} allows the user to give a label (e.g., {\lmttfont b1}, {\lmttfont b2}) to the parts of the code pattern that matched the software-under-injection, and to reuse these parts in the code replacement.


In the second example (\figurename{}~\ref{fig:dsl_example_2}), the MIFS fault type matches an IF construct with its statements (up to 4), and removes them, i.e., the code replacement part of the specification is empty. The specification mixes fragments of Python code (i.e., the {\lmttfont if} construct and {\lmttfont continue} keywords) and DSL directives ({\lmttfont \$EXPR}, {\lmttfont \$BLOCK}). Again, we refined the original fault type from G-SWFIT by leveraging domain knowledge, to inject into more specific targets. We emulate another recurring issue in OpenStack, in which metadata of resources (e.g., the {\lmttfont UUID} of instances) must have been initialized to allow operating on the resource, but a check on the validity of the metadata has been omitted (e.g., the Nova bug \#1096722 \cite{bugopenstack2}). To emulate this real bug, we target {\lmttfont if} constructs that check specific variables (e.g., variables called {\lmttfont node}, which are used throughout the OpenStack Nova codebase) and that skip an operation if the check fails (e.g., by issuing a {\lmttfont continue}).

In the third example (\figurename{}~\ref{fig:dsl_example_3}), the WPF fault type injects an invalid parameter to a function call. The bug specification replaces a {\lmttfont\$CALL} statement with the same {\lmttfont\$CALL} statement, but modifying one of the input parameters. We use again a {\lmttfont tag} to reuse code from the code pattern in the code replacement, by means of the {\lmttfont \#c} syntax after {\lmttfont \$CALL}, i.e., the matched function call is labeled as ``\emph{c}''. We tailored the bug specification to match another recurring issue in OpenStack, in which an external utility (e.g.,  {\lmttfont iptables}, {\lmttfont dnsmasq}, {\lmttfont e2fsck}) is invoked at the host OS level, but with incorrect or missing parameters (e.g., the Nova bug \#732549 \cite{bugopenstack3}). Thus, we target the {\lmttfont utils.execute()} library function (the {\lmttfont name} attribute in {\lmttfont \$CALL}), and look for a string literal ({\lmttfont \$STRING}) among the input parameters of the function, where the string contains the \emph{\-} character used by UNIX utilities to denote parameters. In the code replacement, we inject the same function call, but the string literal (labeled \emph{s}) is wrapped by a function call that corrupts the string with random contents, using the {\lmttfont\$CORRUPT} DSL directive.

In addition to these examples, we have been using the DSL to define several fault models in an industrial context, in cooperation with Huawei Technologies Co. Ltd.. The DSL provided us a fine-grain control over the injections, by combining DSL directives with Python code fragments. Other fault types include: the injection of exceptions within {\lmttfont try} blocks, in order to increase the test coverage of error handlers \cite{joshi2011prefail,marinescu2009lfi}; the injection of {\lmttfont None} values from library function calls, in order to test error handlers in which the returned value is checked by an {\lmttfont IF} construct after the call; the omission of optional input parameters to function calls; the omission of AND/OR clauses in {\lmttfont IF} conditions; wrong or missing initialization of data, such as key-value pair literals in Python dictionaries, using the {\lmttfont\$CORRUPT} directive; high resource consumption (CPU, memory, storage), using the {\lmttfont\$HOG} directive. 
The DSL can be used to inject more complex fault types, by: using regular expressions for specifying search patterns; using the tagging syntax in the \textit{change} block, to change the order of statements in the \textit{into} block; mutating any arithmetic, boolean, and control flow expression of the Python grammar; injecting
algorithmic bugs by removing entire portions of code (e.g., patterns with multiple nested loops and control flow constructs), and by injecting artificial time delays using a {\lmttfont\$TIMEOUT} directive. 
More examples are presented in \S{}~\ref{sec:case_study}.

\section{The \toolname{} workflow}
\label{sec:pycast_overview}

\begin{figure*}
  \centering
  \includegraphics[width=\textwidth]{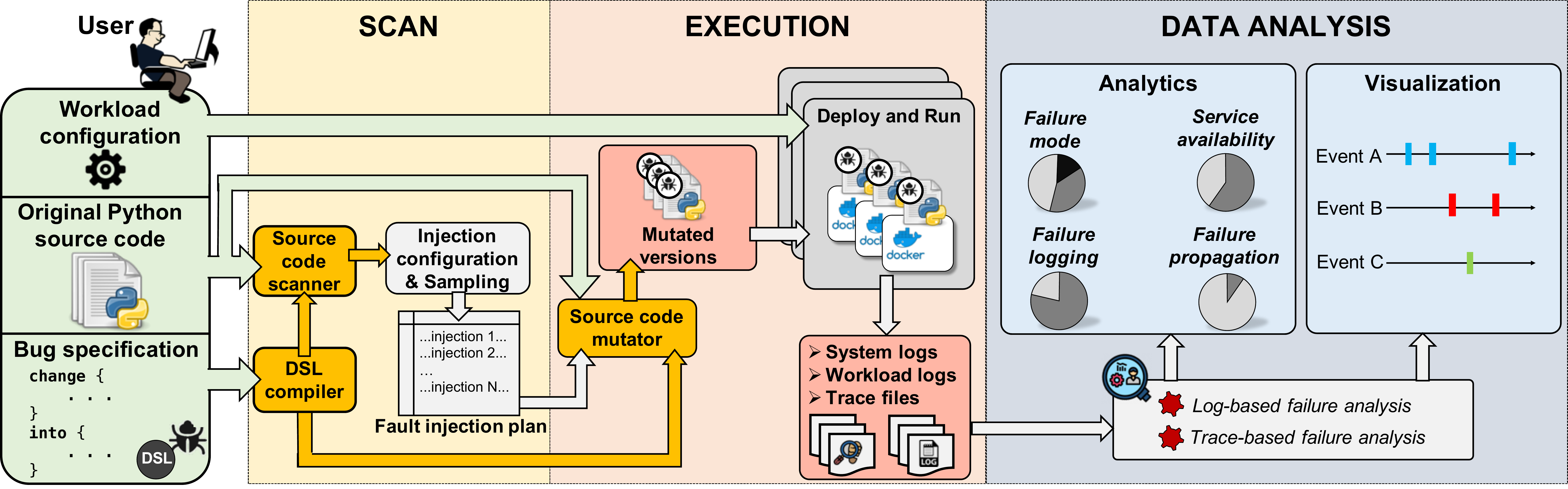}
  \vspace{-20pt} 
\caption{Workflow of the \toolname{} tool.}
  \label{fig:pycast_workflow}
\vspace{-10pt} 
\end{figure*}

\toolname{} provides a complete fault injection workflow, which assists test engineers at applying software fault injection in Python systems. The \toolname{} workflow  generates a set of mutated versions of the target software, according to user-defined bug specifications. These mutated versions are executed in a controlled environment, and further analyzed for drawing insights about the system behavior under failure.
\figurename{}~\ref{fig:pycast_workflow} summarizes the workflow, which consists in a sequence of three main phases, that is, \textit{Scan} (see \S{}~\ref{subsec:pycast_scan_phase}), \textit{Execution} (see \S{}~\ref{subsec:pycast_execution_phase}), and \textit{Data Analysis} (see \S{}~\ref{subsec:pycast_data_analysis_phase} and \S{}~\ref{subsec:pycast_additional_features}). The following sub-sections provide details for each phase. 

\subsection{Scan}
\label{subsec:pycast_scan_phase}

In the \textit{Scan} phase, the user interacts with the \toolname{} tool to define the \emph{fault injection plan}, which is the set of fault injection experiments to be run. Each experiment specifies a fault to be injected. 
\toolname{} takes in input the source code of the target software, and the bug specification described by using our DSL (\sectionautorefname{}~\ref{sec:pycast_dsl}). 
The fault model is stored in a JSON file, and users can save and import fault models of previous fault injection campaigns. \toolname{} provides pre-defined fault models based on previous fault injection studies (section~\ref{sec:related}).

The \textit{Scan} phase identifies \textbf{fault injection points} in the software, i.e., a statement (or group of statements) in the source code where \toolname{} can inject the software bug according to the user-defined specification. \toolname{} looks for arithmetic/boolean expressions, method and function calls, variable initializations, and other kinds of statements. 

\toolname{} processes the target code using its Abstract Syntax Tree (AST) representation, which is commonly by program analyzers to represent the structure of a piece of code. 
The \textit{DSL compiler} component takes the bug specification written using the DSL and generates a meta-model, which consists of a small AST that reflects the structure of the code in the code pattern. The meta-model will be used by the \textit{source code scanner}, which visits the program's AST to find matches against the code pattern  (i.e., portions of the program's AST that match the AST of the meta-model). The meta-model is also used by the \textit{source code mutator} to generate fault-injected versions of the program (see \S{}~\ref{subsec:pycast_execution_phase}). 

After obtaining a set of fault injection points, the user can select a subset of such locations according to their needs. For example, the user may want to perform experiments only for a specific component (e.g., class or file); the user may want to inject a sample of randomly-chosen faults (e.g., to enforce a limit on the number of experiments); or, the user can inject faults in all of the injection points. The set of injections defines the fault injection plan, which is used in the \textit{Execution} phase.


In this paper, the proposed DSL is tailored for the Python language. It is possible to define a similar DSL to support other languages, such as C/C++ and Java. Several of the bug patterns for Python could be re-used (i.e., patterns not involving special Python syntax). The porting would mostly affect the DSL compiler and the source code scanner and mutator.

\subsection{Execution}
\label{subsec:pycast_execution_phase}

In this phase, \toolname{} iterates over the fault injection plan. In each experiment, the \emph{original} Python source code is transformed into a \emph{mutated} version, which is identical to the original except for a few mutated statements. The mutation emulates a residual bug in the software. For example, to inject a wrong parameter bug in a method call, \toolname{} modifies the method call statement by replacing it with a call to the same method but with different or corrupted input parameters; to emulate an omission by the developer, \toolname{} deletes the method call in the mutated version. The set of mutated versions are the \emph{faultload} that will be executed in the experiments. 
At the end of every experiment, \toolname{} collects logs from the target system for data analysis (\S{}~\ref{subsec:pycast_data_analysis_phase}).


The user also configures a \emph{workload}, i.e., a set of directives to exercise the target software during the experiments. The workload emulates the operating conditions of the system and triggers the injected fault. Moreover, the workload serves to detect service failures and recovery abilities, e.g., by looking for crashes and timeouts of the workload (e.g., due to stalled service calls), or by performing consistency checks with test assertions on the outputs of the workload (e.g., after a resource has been modified by the workload, the behavior of the system should reflect the new state of the resource).

The user defines the workload by providing command-line directives. For example, the user can use UNIX shell commands to start the target software, e.g., to launch a UNIX daemon such as a network server. 
Command-line directives can be used both to invoke the command-line interface of the target Python program or to indirectly launch the software by running automated test scripts. These scripts can be uploaded by the user along with the target Python source code (\figurename{}~\ref{fig:pycast_workflow}). Additionally, the user can specify command-line directives to launch workload generator tools, such as HTTP and RPC traffic generators, which in turn exercise the target software. 

\toolname{} runs the fault injection experiments within a \emph{container-based} experimental environment, by using the Docker virtualization system \cite{docker}. The tool first creates a container image, in which it copies the Python source code uploaded by the user. The user can customize the container image by adding configuration directives in \emph{Dockerfile} format \cite{dockerfile_ref}, such as, to install within the container external dependencies to run the Python software under test (e.g., using the \emph{pip} command), and to install external tools (e.g., HTTP and RPC traffic generators). Then, for each fault to be injected, \toolname{} deploys a new container, by copying into it the mutated source code with the fault, and runs the workload directives defined by the user. The experiment ends when the workload completes, or when a user-defined timeout expires. Finally, \toolname{} cleans-up the experimental environment by deallocating the container. In this way, the tool can also clean-up any resource leaked or corrupted because of the injected fault (e.g., stale processes or files). Using containers also allows the tool to run several parallel experiments on independent sandboxes, to take advantage of multi-core CPUs. \toolname{} tunes the number of parallel experiments according to run at most $N-1$ parallel containers at the same time, where $N$ is the number CPU cores in the host system \cite{winter2015no}. To avoid interferences in memory and I/O bandwidth, the tool further reduces the number of parallel containers if it hits a threshold for memory and I/O utilization.

\toolname{} can enable and disable the injected faulty code at any time during the execution of the target software. 
The mutated source code retains a copy of the original statements of the fault injection point, similarly to the EDFI fault injection tool \cite{giuffrida2013edfi}: \toolname{} mutates the source code by inserting an {\lmttfont IF ... ELSE ...} construct, where the two branches include respectively the original statements and the faulty ones. Then, the tool can control which of the two branches to execute, by writing a control variable (a ``\emph{trigger}'') allocated in a shared memory area between the tool and the target software. 
This ability enables additional analyses of the effects of failures and recovery. The tool executes the workload for two times (``\emph{rounds}''), without restarting the target program between the two executions. In the first round, the injected fault is enabled, so that it infects the target software with error states, possibly causing service failures. The workload is executed again in the second round, but the injected fault is disabled. Of course, if the target program fails and is unable to recover, the second workload execution will fail. The second round allows us to analyze the \emph{scope} of the error states \cite{yoshimura2013using,sugimoto2019short}. In the best case, the error state is confined to service requests that were issued during the first round, and the requests during the second round are not affected by any error (e.g., the target software recovers a correct state with a restart). In the worst case, the error states are persistent even after that the faulty code is disabled, causing further failures during the second round. This analysis provides additional feedback to the user about the failure behavior of the target software.

During the experiments, \toolname{} saves the output of the target program (\emph{stdout}, \emph{stderr}) and the output of the workload directives (e.g., the commands for launching a workload generator, which reports service failures). Moreover, the tool can be configured to save log files that may be generated by the target software or by the workload. These outputs and logs are analyzed in the last phase of the \toolname{} workflow (\emph{data analysis}), as discussed in the following.

\subsection{Data Analysis}
\label{subsec:pycast_data_analysis_phase}


The \emph{data analysis} evaluates the target software in terms of service failures, logging, and recovery. 
\toolname{} classifies the experiments into a set of \textbf{failure modes}, which include the crash and the timeout of the target software, and user-defined failure modes. The user can specify patterns (e.g., using keywords and regex) that the tool will look for among the outputs and the logs produced by the experiments. For example, failure modes can include failures of the workload (e.g., the workload stops due to a service API exception) and of the target software (e.g., the software detects an error state with an internal assertion, and reports it with a high-severity log message). The tool reports the statistical distribution of failure modes. The user can drill-down the individual classes of failures, to further inspect logs of experiments in that class. The user can also drill-down with respect to fault types and injected components, to identify the critical areas (e.g., components that are most prone to failures) where failure mitigations are most needed.

\toolname{} can analyze failures with respect to workload rounds. It computes a \emph{service availability} metric, i.e., the percentage of experiments in which the software was (un)available in the second round of execution (injected fault disabled), because of error states generated during the first round (injection fault enabled) that persisted and were not recovered. These cases deserve a deeper analysis, e.g., to identify resource leaks that may occur in error handling paths, and that may cause more failures over time \cite{huang1995software,grottke2007:fightingbugs}.

\subsection{Advanced Features}
\label{subsec:pycast_additional_features}

\toolname{} includes more, optional features for deeper analysis of the large amounts of data produced by fault injection experiments. We briefly report here on these features.

\vspace{1pt}
\noindent
$\blacksquare$ \textbf{Coverage analysis}. To reduce the time needed to run the fault injection experiments, \toolname{} performs a preliminary analysis to avoid injecting faults in \emph{program paths} that are not \emph{covered} by the workload.  
Most likely, the workload will not cover all of the paths in the program, and injecting into non-covered paths causes a waste of time since the fault would not cause any effect. 
Before executing the experiments, \toolname{} conducts a \textit{coverage analysis}, by running a ``fault-free'' execution (i.e., no fault injected) using the same workload that will be used for the experiments. It generates coverage information by adding logging statements at every fault injection point in the target program discovered by the \emph{scan} phase (see \S{}~\ref{subsec:pycast_scan_phase}). After the fault-free run, \toolname{} generates a \textit{reduced} fault injection plan, by only including the covered fault locations.

\vspace{1pt}
\noindent
$\blacksquare$ \textbf{Failure logging}. \toolname{} checks whether the target system can detect error states and report diagnostic information on \emph{log files}. 
The tool computes a \emph{failure logging} metric, i.e., the percentage of experiments in which the target software both experienced a workload failure and logged at least one error message. Failures and error logs are identified with user-provided keywords and regex. This metric gives feedback about the logging abilities, and non-logged failures are opportunities for improving telemetry. An example of this analysis can be found in a previous study \cite{cotroneo2019bad}.

\vspace{1pt}
\noindent
$\blacksquare$ \textbf{Failure propagation}. \toolname{} checks if the fault in the injected component propagated across other components. 
The tool computes a \emph{failure propagation} metric, i.e., the percentage of injected faults that impacted on more than one component. This metric is applicable for larger software with a component-based architecture, where each sub-system generates a distinct log file, or where logs of the sub-systems can be separated with keywords and regex. The user configures a list of sub-systems, their source code files (e.g., a sub-folder of the source code), and their log files or patterns. The experiments that exhibit propagation are worth further investigation, e.g., to develop more robust interfaces between sub-systems to prevent the propagation and make recovery easier. Examples of this analysis can be found in previous papers \cite{cotroneo2019bad,cotroneo2019enhancing}.

\vspace{1pt}
\noindent
$\blacksquare$ \textbf{Failure visualization}. \toolname{} provides a graphical representation of an experiment, to help the user to understand what happened during a failure. The tool instruments selected RPC APIs in the target software, and records their invocations during the experiment using the \emph{Zipkin} distributed tracing framework \cite{zipkin}. These API calls are visualized as \emph{events} on timelines as interactive plots. An example of visualization can be found in a previous study \cite{cotroneo2019failviz}.

\section{Case Study}
\label{sec:case_study}

\begin{table}[!t]
\caption{Injected fault types.}
\label{tab:fault_category}
\centering
\begin{tabular}{
>{\centering\arraybackslash}p{2cm} >{\centering\arraybackslash}p{3.4cm} >{\centering\arraybackslash}p{2.1cm}
}
\toprule
\textbf{Fault Category} & \textbf{Injection Target} & \textbf{Examples of Injections}  \\ 
\midrule
Failures when calling external library APIs &
API calls to the {\lmttfont urllib} and  {\lmttfont os} Python modules &
Exceptions,  {\lmttfont None} objects, omitted call, wrong call\\
\midrule
Wrong inputs in Python-etcd API &
{\lmttfont set(key, val)}, {\lmttfont get(key)}, {\lmttfont test\_and\_set(key, val, old)}, ... &
String corruptions,  {\lmttfont None} values, negative integers\\
\midrule
Resource management bugs &
{\lmttfont set(key, val)}, {\lmttfont get(key)}, {\lmttfont test\_and\_set(key, val, old)}, ... &
Hog threads inside methods of Python-etcd\\
\bottomrule
\end{tabular}\vspace{-15pt}
\end{table}

We present an application of \toolname{} in the context of \emph{Python-etcd} \cite{python-etdc}, which is a library that provides Python bindings for the \emph{etcd} distributed key-value store \cite{etcd_server}. 
%
%
Huawei uses \emph{Python-etcd} in their systems and asked for three fault classes to be evaluated using our fault injection tool (Table~\ref{tab:fault_category}): (i) call failures when invoking APIs from external libraries (wrong response, timeouts, etc.), (ii) wrong inputs to the \emph{Python-etcd} APIs, and (iii) resource management faults. We implemented these fault types using the \toolname{} DSL language.

%

We performed three fault injection campaigns on \emph{Python-etcd} version 0.4.5. The workload used deploys the \emph{etcd} server, and it uploads and queries several key-value pairs of a different kind (e.g., with directories, sub-keys, TTL, etc.) that we derived from \emph{Python-etcd}'s integration tests.  
In the following subsections, we present the injected fault types and analyze failure modes using \toolname{}. 

\subsection{Errors from external APIs}

In the first campaign of experiments, we injected faults at method calls in Python-etcd external modules, targeting the methods of {\lmttfont urllib} (a Python package for working with URLs) and from {\lmttfont os} (e.g., Python methods for file I/O). 
The injected fault types include:

\begin{itemize}

    \item \textbf{Throw Exception}: The {\lmttfont raise} of the exception on a method call, according to pre-defined, per-API list of exceptions (e.g., {\lmttfont ConnectTimeoutError});
    
    \item \textbf{Missing Function Call}: A method call is entirely omitted (e.g., replaced with the python statement {\lmttfont pass});
    
    \item \textbf{Missing Parameters}: A method call is invoked with omitted parameters (e.g., the method uses a default parameter instead of the correct one). 
    
\end{itemize}

For this faultload, \toolname{} identified \numprint{26} points where to inject faults. In \numprint{13} cases, the workload covered the injected faulty code. We found failures in \numprint{12} experiments.

\vspace{0.8pt}
\noindent
$\rhd$
\textbf{Reconnection failure}. In half of the cases, we found failures in both rounds of execution, as denoted by the \emph{service availability} metric. The experiments did not complete within the timeout, and {\lmttfont etcd} was unable to reconnect even after the fault removal. We found that the {\lmttfont etcd} server was unable to bind to a TCP/IP port. Thus, restarting {\lmttfont etcd} does not suffice to recover from the fault, but the port needs to be explicitly freed. 
We need additional exception handlers to catch exceptions caused by network connections, such as time-outs.

\vspace{0.8pt}
\noindent
$\rhd$
\textbf{Critical errors about {\lmttfont 'member has already been bootstrapped'}}. In a few experiments, Python-etcd was unable to perform operations on {\lmttfont etcd} in the first round, due to an inconsistent state of the server caused by the fault. To recover from this failure, the system needs a more elaborated exception handling: it should explicitly remove the affected member by using the dynamic configuration API of {\lmttfont etcd}, and it should restart {\lmttfont etcd} by reverting to a previous consistent state.

\vspace{0.8pt}
\noindent
$\rhd$
\textbf{Client process crash due to an exception}. In the remaining cases, the client process crashed during the first round due to an unhandled exception. Moreover, the system was not available after disabling the fault. In these cases, Python-etcd should provide exception handlers to catch these exceptions or to raise another kind of exception (such as \textit{EtcdException}) to be managed by Python-etcd client process.

\subsection{Wrong Inputs}

In the second campaign of fault injection experiments, we injected faults in input parameters of Python-etcd API methods. We configured \toolname{} with fault types for injecting corrupted inputs, such as strings with random characters, None object references, negative integers, etc. 
For example, let us consider the {\lmttfont method test\_and\_set(key, value, old\_value)} taking in input three parameters: A fault consists in injecting a corrupted input in the first parameter (string type) by randomly replacing the characters of the string. 

The \toolname{} tool identified \numprint{66} locations where to inject these faults.
In all of the cases, the injected faulty code was covered by the workload, and in \numprint{29}  experiments we found the following failures in the first round of execution:

\vspace{0.8pt}
\noindent
$\rhd$ \textbf{AttributeError: ’NoneType’ object has no attribute ’startswith’}. This failure is due to an issue of Python-etcd. It happens when the tool injects a {\lmttfont None} value instead of a string (e.g., a \emph{key} string). Python-etcd does not check whether the input strings are valid. Therefore, when a {\lmttfont None} value is passed in input, Python-etcd uses the {\lmttfont startswith} attribute on a {\lmttfont None} reference. To avoid this failure, Python-etcd should sanitize null strings in inputs.

\vspace{0.8pt}
\noindent
$\rhd$ \textbf{EtcdKeyNotFound exception}. This failure happens when a wrong key or value is injected. In this case, the workload failed because it is not able to find the expected key or value in the {\lmttfont etcd} datastore. The caller (in this case, the workload) needs to get/set the correct keys and values. Thus, the Python-etcd client should handle these exceptions.

\vspace{0.8pt}
\noindent
$\rhd$ \textbf{EtcdException: Bad response: 400 Bad Request}. This failure happens when \toolname{} injects a wrong key or value that is not valid (e.g., a non-ASCII string). When this value is passed to {\lmttfont etcd}, the server rejects the request with the \emph{HTTP Error 400 Bad Request}. Python-etcd should be fixed to check and sanitize non-ASCII strings.

\subsection{Resource Management Bugs}

In the last campaign of experiments, we injected CPU hogs to overload Python-etcd. We used \toolname{} for injecting stale threads that generate a high CPU load. We targeted the same methods of the second campaign of experiments, by injecting a resource hog after the method call.  
The tool found \numprint{37} injectable locations, and the faulty code was always covered during the workload execution. 
In \numprint{14} experiments, the system experienced a service failure in the first round of execution. Most of these failures forced a process termination with the exception \emph{``UnboundLocalError: local variable ... referenced before assignment''}. In other cases, the workload also failed because of inconsistent values read from the {\lmttfont etcd} datastore. 
The high CPU usage triggered race conditions in Python-etcd, and in the Python interpreter itself. Since it is hard to find and to fix these issues, the failure should be mitigated, by cleaning-up stale threads that may cause high CPU consumption. This should be pursued by monitoring at run-time the CPU utilization of Python processes, and by killing or restarting stale threads if CPU utilization is too high. 

\subsection{Performance evaluation}


\toolname{} can quickly inject faults even for large projects since the scan and mutation can be parallelized across several CPUs (it is an ``embarrassingly parallel`` task). It took less than one minute to scan and mutate \textit{Python-etcd} on an 8-core Intel Xeon with 16 GB RAM. 
We also evaluated performance on the OpenStack project, by targeting the three most important modules (Nova, Neutron, and Cinder) accounting for about 400K lines of Python code. Using the same hardware, \toolname{} takes about 20 min to identify 17488 injectable locations using 120 different DSL patterns, which is reasonable for practical purposes given the large size of this project. 
The duration of the \textit{execution} phase is beyond the control of our tool since it depends on the time to deploy the target system and run the workload. It took between 10s and 120s (worst case of a ``hang'' failure) to run a single experiment on \textit{Python-etcd}, and about 30 min to run all of the tests of this section. For OpenStack, an experiment takes several tens of minutes, since it is a complex system that deploys VMs, loads large storage volumes, initializes databases, etc. We were able to execute experiments on OpenStack through nightly parallelized runs.

\section{Conclusion}
\label{sec:conclusion}

\toolname{} is  designed to be programmable and highly usable, by performing fault injection campaigns with customized faultloads in Python software. 
%
%

The analysis of results pointed out several failure modes, which were acknowledged as valid threats by our industrial partners. The programmability of the tool through a DSL was useful to easily and quickly customize fault injections to comply with the fault classes requested by the company, based on their internal software resiliency requirements. We discussed in the paper potential strategies to mitigate the failure modes.

We plan to extend the tool with more features for failure analysis and to use it as a basis for research on software fault tolerance strategies in modern applications, such as cloud software.


\section*{Acknowledgments}
This work has been done in the framework of the R\&D project of the multiregional investment programme "REINForce: REsearch to INspire the Future” (CDS000609) with Hitachi Rail S.p.A., supported by the Italian Ministry for Economic Development (MISE) through the Invitalia S.p.A. agency.

\IEEEtriggeratref{57}

\Urlmuskip=0mu plus 1mu\relax
\clearpage
\bibliographystyle{IEEEtran}
\bibliography{references-fi,references-misc}

\end{document}